\documentclass[aps,prb,reprint,showpacs,superscriptaddress]{revtex4-1}
\usepackage{graphicx}
\usepackage{dcolumn}
\usepackage{bm}
\usepackage{amssymb}
\usepackage{amsmath}
\usepackage{subfigure}
\usepackage{epstopdf}
\usepackage{float}
\usepackage{mathrsfs}
\usepackage{xcolor}
\begin{document}

\title{Current-driven Dynamics of Magnetic Skyrmion in Chiral Ferromagnetic Film with Spatially Modulated Dzyaloshinskii-Moriya Interaction}
\author{Liping Zhou}
\thanks{These two authors contributed equally}
\affiliation{School of Physics, Nankai University, Tianjin 300071, China}
\author{Ren Qin}
\thanks{These two authors contributed equally}
\affiliation{School of Physics, Nankai University, Tianjin 300071, China}
\author{Ya-qing Zheng}
\affiliation{School of Physics, Nankai University, Tianjin 300071, China}
\author{Yong Wang}
\email{yongwang@nankai.edu.cn}
\affiliation{School of Physics, Nankai University, Tianjin 300071, China} 

\begin{abstract}
The dynamics of magnetic skyrmion driven by spin-polarized current is theoretically studied in the chiral ferromagnetic film with spatially modulated Dzyaloshinskii-Moriya interaction. Three cases including linear, sinusoidal, and periodic rectangular modulations have been considered, where the increase, decrease, and the periodic modification of the size and velocity of the skyrmion have been observed in the microscopic simulations. These phenomena are well explained by the Thiele equation, where an effective force on the skyrmion is induced by the inhomogeneous Dzyaloshinskii-Moriya interaction. The results here suggest that the dynamics of skyrmion can be manipulated by artificially tuning the Dzyaloshinskii-Moriya interaction in chiral ferromagnetic film with material engineering methods, which will be useful to design skyrmion-based spintronics devices.
\end{abstract}

\maketitle

\section{Introduction}
In recent years, the discovery of magnetic skyrmion in chiral ferromagnetic materials has triggered extensive research activities,\cite{JMMM1994,PRL2001,Nat2006,Sci2009,Nat2010,NM2010,NP2011,NP2012,PRL2009,PRL2011-1,PRL2011-2,Sci2013,NN2013-1,NN2013-2,PRL2013,NC2014,
Sci2015,NP2017-1,NP2017-2,NN2017} mainly due to its potential applications in spintronics devices.\cite{NN2013-3,NRM1,NRM2,Proc,NatEle2018} This type of topological defect is stabilized by the finite Dzyaloshinskii-Moriya interaction (DMI),\cite{Dzya,Moriya} which origins from the broken inversion symmetry in the materials. The dynamics of the skyrmion can be flexibly manipulated by spin-polarized current,\cite{NN2013-1,NN2013-2} and the trajectory of its center can be simply described by the Thiele equation.\cite{NN2013-3} It shows that a transverse component of the velocity will be generated by an effective force on the skyrmion, \emph{i.e.} ``the skyrmion Hall effect",\cite{NN2013-2} which has been confirmed in recent experimental observations.\cite{NP2017-1,NP2017-2} 

Meanwhile, the physical factors influencing the DMI in chiral ferromagnetic materials have been carefully investigated.
\cite{PRB2015,SR2015,PRB2016,SR2016,PRL2017-1,PRL2017-2,NL2018,PRB2018,Nanoscale2018,SR2018} For instance, it has been shown that the bulk DMI can be modified by the chemical composition,\cite{PRB2015} carrier density and the strain,\cite{SR2015}, band filling,\cite{PRL2017-2} etc.  The interfacial DMI in the multilayer thin films, on the other hand, can be tuned by the thickness of certian layer,\cite{PRB2016,Nanoscale2018,SR2018} oxygen coverage,\cite{SR2016} ion irradiation,\cite{PRL2017-1} etc. Furthermore, the effect of electric field has also been studied,\cite{NL2018,SR2018} where 130\% variation of DMI in Ta/FeCoB/TaO$_{x}$ trilayer has been demonstrated experimentally.\cite{NL2018} These developments thereby lay the foundation to control the dynamics of magnetic skyrmion by artificially modulate the DMI in chiral ferromagnetic materials.

By far, the current-driven dynamics of magnetic skyrmion has mainly been studied in chiral ferromagnetic materials with homogeneous DMI. In this paper, we consider a chiral ferromagnetic film with spatially modulated DMI, and show that the trajectory of the skyrmion will be drastically modified due to the inhomogeneous distribution of DMI, which can be well understood by analyzing the Thiele equation. The results here will be beneficial to design skyrmion-based devices by controlling the DMI with material engineering techniques.

The rest of paper is organized as follows. In Section II, we describe the theoretical model of the current-driven skyrmion with inhomogeneous DMI, and give the corresponding Thiele equation. In section III, we perform micromagnetic simulations with linearly or periodically modulated DMI, and compare the results with Thiele equation. The conclusive remarks of our work will be given in Section IV.

\section{Theoretical Model}
\begin{figure}[!ht]
  \centering
  \includegraphics[width=\linewidth]{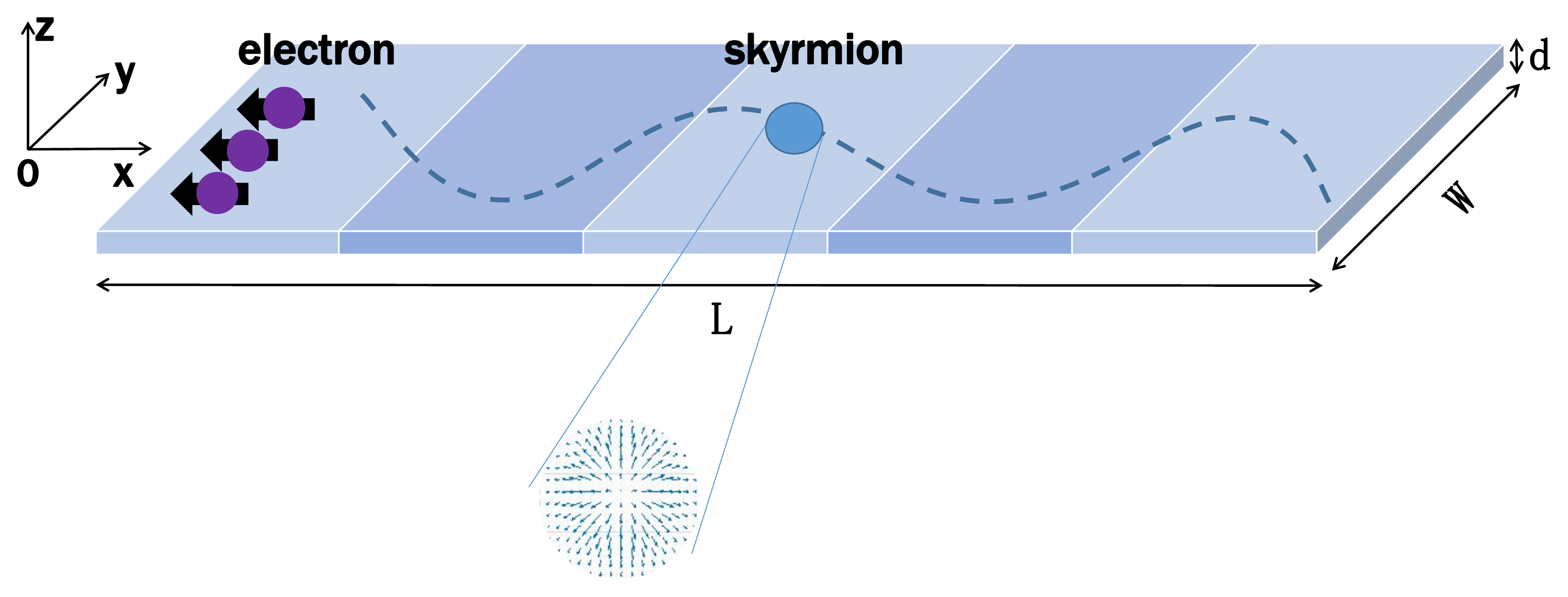}
  \caption{(Color online) Schematic diagram of the system under consideration. A chiral ferromagnetic film with spatially modulated Dzyaloshinskii-Moriya interaction is placed in the $x$-$y$ plane. A magnetic skyrmion driven by spin-polarized current is moving in the film. $L$ : length; $W$ : width; $w$ : thickness. }
  \label{Fig1}
\end{figure}

As shown in Fig.~\ref{Fig1}, the model system is a chiral ferromagnetic film with length $L$, width $W$, and thickness $d$. The main interactions among the magnetic moments here include the Heisenberg exchange interaction, DMI, and the perpendicular magnetic anisotropy, which thus give the energy density of the film as\cite{NN2013-1,NN2013-2,NN2013-3}
\begin{eqnarray}
\mathcal{E}[\mathbf{m}]=A(\nabla\mathbf{m})^{2}+\mathcal{E}_{DM}[\mathbf{m}]-K_{u}(\mathbf{e}_{z}\cdot\mathbf{m})^{2}.\label{Ham}
\end{eqnarray}
Here, $\mathbf{m}(\mathbf{r})=\mathbf{M}(\mathbf{r})/M_{s}$ is the normalized magnetization distribution, and $M_{s}$ is the saturation magnetization; $A$ is the exchange stiffness; $K_{u}$ is the perpendicular magnetic anisotropy coefficient, and $\mathbf{e}_{z}$ is the anisotropic axis along $z$-direction. $\mathcal{E}_{DM}[\mathbf{m}]$ represents the energy density due to DMI interaction, where $\mathcal{E}_{DM}[\mathbf{m}]=D(m_{z}\nabla\cdot\mathbf{m}-\mathbf{m}\cdot\nabla m_{z})$ for interfacial DMI and $\mathcal{E}_{DM}[\mathbf{m}]=D\mathbf{m}\cdot\nabla\times\mathbf{m}$ for bulk DMI.\cite{NN2013-1,NN2013-2,NN2013-3} The coefficient $D$ determines the strength of DMI and is assumed to be spatially inhomogeneous. 

A magnetic skyrmion will be prepared near the upper-left corner of the film, and a spin-polarized current perpendicular to the film will be applied. The amplitude, degree and direction of spin polarization of the applied current are denoted by $I$, $p$, $\mathbf{m}_{p}$, respectively. Then the magnetization dynamics of the film will be described by the Landau-Lifshitz-Gilbert-Slonczewski (LLGS) equation
\begin{eqnarray}
\frac{d\mathbf{m}}{dt} = -\gamma_{0}\mathbf{m}\times\mathbf{H}_{eff} + \alpha\mathbf{m}\times\frac{d\mathbf{m}}{dt}
  -\beta\mathbf{m}\times(\mathbf{m}\times\mathbf{m}_{p}).\nonumber\\\label{LLGS}
\end{eqnarray} 
Here, $\gamma_{0}$ is the gyromagnetic ratio; $\mathbf{H}_{eff}=-\frac{1}{M_{s}}\nabla_{\mathbf{m}}\mathcal{E}[\mathbf{m}]$ is the effective magnetic field; $\alpha$ is the Gilbert damping coefficient. The last term in Eq.~(\ref{LLGS}) is the Slonczewski spin transfer torque with the amplitude $\beta=\frac{\gamma_{0} \hbar Ip}{2dM_{s}e}$, where $\hbar$ and $e$ are the reduced Planck constant and elementary charge, respectively. The solution of Eq.~(\ref{LLGS}) will give the current-driven dynamics of the skyrmion. Before going to the detailed micromagnetic simulations, we will first analyze the motion of skyrmion based on the Thiele equation. \cite{Book}  

If the spatial variation of the DMI is small enough in the scale of the skyrmion size, we can approximately describe the structure of the skyrmion with fixed DMI strength $D$ at its center. Its trajectory then can be described by the Thiele equation (Appendix A)
\begin{eqnarray}
\hat{\mathcal{G}}\bm{v}-\alpha\hat{\mathcal{D}}\bm{v}-\mathbf{F}-\mathbf{F}^{s}=0.\label{Thiele}
\end{eqnarray} 
Here, $\bm{v}$ is the drift velocity of the skyrmion center; $\hat{\mathcal{G}}$ and $\hat{\mathcal{D}}$ denote the gyromagnetic coupling tensor and dissipative force tensor; $\mathbf{F}$ and $\mathbf{F}^{s}$ are the effective forces acting on the skyrmion due to the inhomogeneous energy density and the spin transfer torque, respectively. They are explicitly related to the magnetization distribution $\mathbf{m}(\mathbf{r})$ as
\begin{eqnarray}
\hat{\mathcal{G}}_{\xi'\xi}&=&\int d\mathbf{r}(\partial_{\xi'}\mathbf{m}\times\partial_{\xi}\mathbf{m})\cdot\mathbf{m},\label{Gvector}\\
\hat{\mathcal{D}}_{\xi'\xi}&=&\int d\mathbf{r}\partial_{\xi'}\mathbf{m}\cdot\partial_{\xi}\mathbf{m},\label{Dtensor}\\
\mathbf{F}_{\xi}&=&-\frac{\gamma_{0}}{M_{s}d}\int d\mathbf{r}\partial_{\xi}\mathcal{E}[\mathbf{m}(\mathbf{r})],\label{ForceU}\\
\mathbf{F}_{\xi}^{s}&=&\beta\int d\mathbf{r}(\partial_{\xi}\mathbf{m}\times\mathbf{m})\cdot\mathbf{m}_{p}.\label{ForceS}
\end{eqnarray} 
The variables $\xi,\xi'$ here denote the spatial coordinates $x,y$. Then the velocity $\bm{v}$ of the skyrmion will be obtained from the Thiele equation as
\begin{eqnarray}
v_{x}&=&\frac{\mathcal{G}(\mathbf{F}_{y}+\mathbf{F}_{y}^{s})-\alpha\mathcal{D}(\mathbf{F}_{x}+\mathbf{F}_{x}^{s})}{\mathcal{G}^{2}+\alpha^{2}\mathcal{D}^{2}},\label{vx}\\
v_{y}&=&-\frac{\mathcal{G}(\mathbf{F}_{x}+\mathbf{F}_{x}^{s})+\alpha\mathcal{D}(\mathbf{F}_{y}+\mathbf{F}_{y}^{s})}{\mathcal{G}^{2}+\alpha^{2}\mathcal{D}^{2}},\label{vy}
\end{eqnarray}
which then gives the Hall angle $\vartheta_{H}=\arctan(v_{y}/v_{x})$.

\begin{figure}[!ht]
  \centering
  \includegraphics[width=\linewidth]{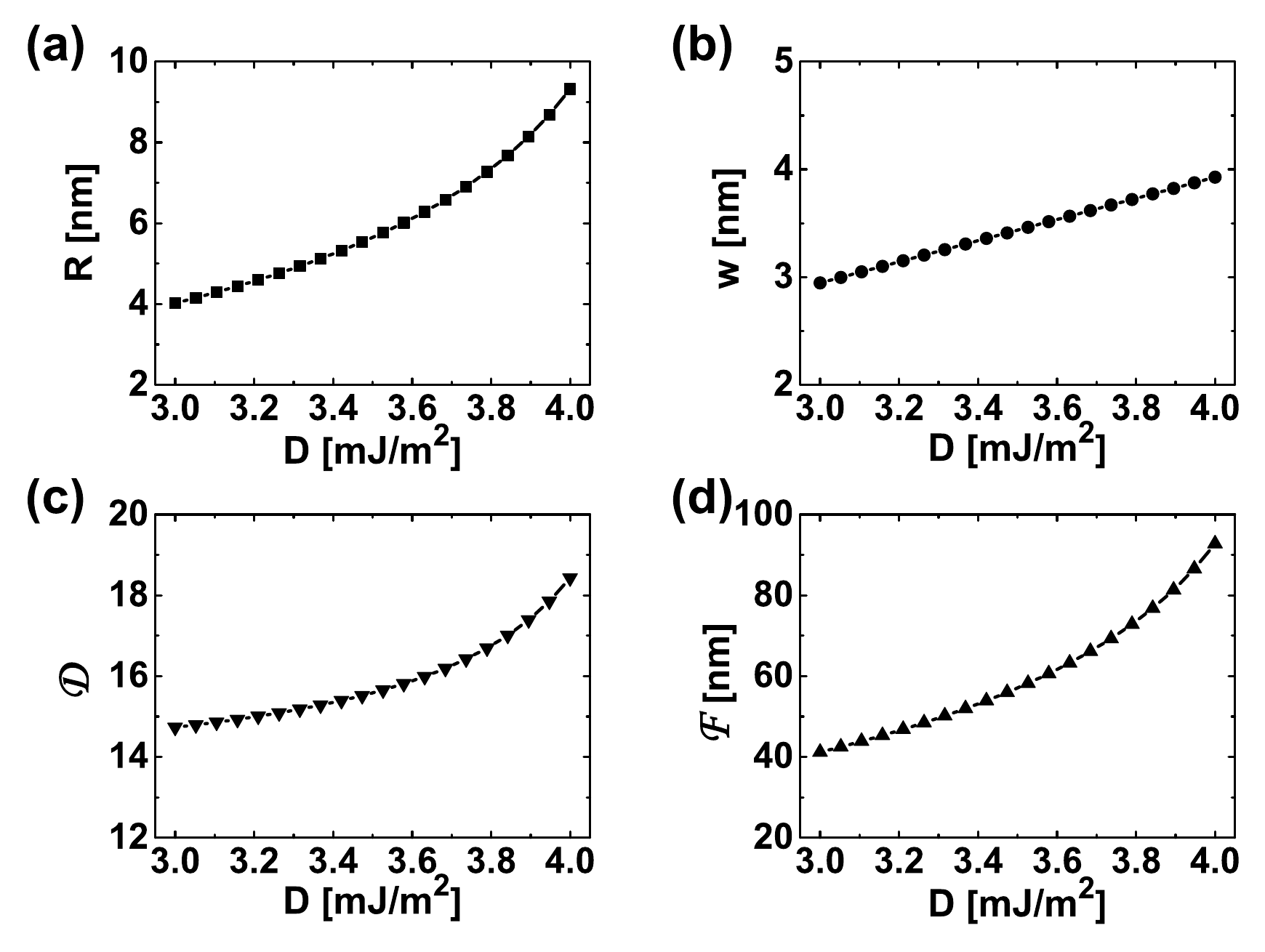}
  \caption{(a)(b) The dependence the radius $R$ and width $w$ of the skyrmion on the DMI strength $D$ according to the model in Ref.~\onlinecite{CP2018}. (c)(d) The integrals $\mathcal{D}$ defined in Eq.~(\ref{D}) and $\mathcal{F}$ defined in Eq.~(\ref{ForceD}) as functions of $D$. We have set $A=15$~pJ/m and $K_{u}=0.8$~MJ/m$^{3}$ in the calculations. }
  \label{Fig2}
\end{figure}

\begin{figure*}[!htp]
  \centering
  \includegraphics[width=\linewidth]{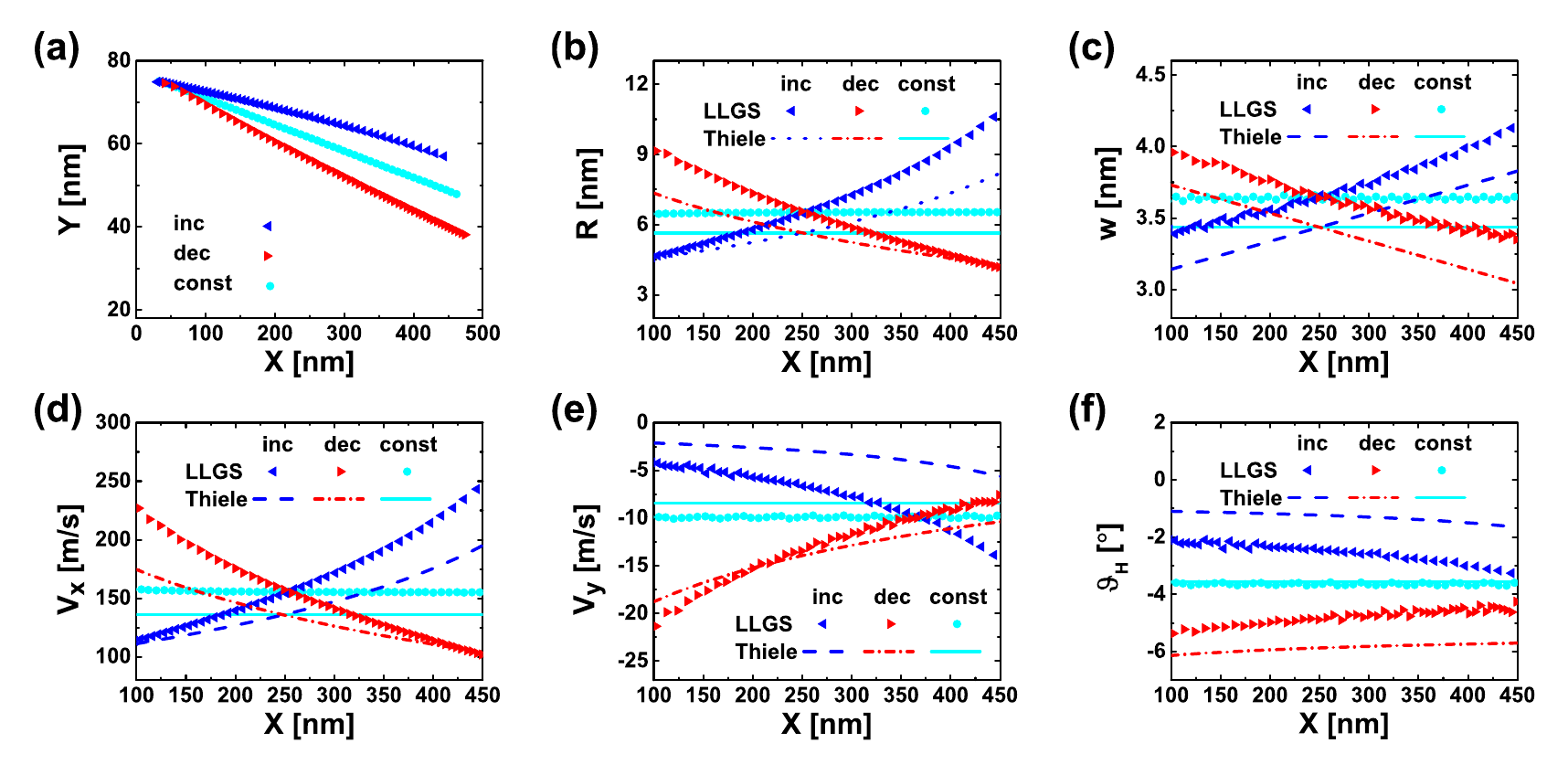}
  \caption{(Color online)(a) The trajectory of the skyrmion in the $x$-$y$ plane of the film; (b) the location-dependent skyrmion radius $R$; (c) the location-dependent skyrmion width $w$; (d)(e) the location-dependent velocity components $v_{x}$ and $v_{y}$ of the skyrmion; (f) the location-dependent Hall angle $\vartheta_{H}$ of the skyrmion. LLGS: micromagnetic simulation based on LLGS equation (\ref{LLGS}); Thiele : theoretical results based on Thiele equation. inc : $D$ increases from $3$~mJ/m$^{2}$ to $4$~mJ/m$^{2}$ for $x\in[0,500]$~nm; dec : $D$ decreases from $4$~mJ/m$^{2}$ to $3$~mJ/m$^{2}$ for $x\in[0,500]$~nm; const : $D$ takes constant value $3.5$~mJ/m$^{2}$ for comparison. }
  \label{Fig3}
\end{figure*}

The interfacial and bulk DMI will result in N\'{e}el-type and Bloch-type skyrmion, respectively. In the polar coordinates taking the skyrmion center as the origin, the magnetization configuration of the skyrmion can be expressed as\cite{NN2013-3} $\mathbf{m}(r,\varphi)=(\sin\Theta(r)\cos\Phi(\varphi),\sin\Theta(r)\sin\Phi(\varphi),\cos\Theta(r))$, where $\Phi=\varphi+\gamma$. Depending on $D>0$ or $D<0$, the helicity $\gamma$ will be $0$ or $\pi$ for the N\'{e}el-type skyrmion, and will be $\pi/2$ or $-\pi/2$ for the Bloch-type skyrmion (Appendix B). Further calculations (Appendix C) give the gyromagnetic coupling tensor 
$ \hat{\mathcal{G}}=\left(\begin{array}{cc} 0 & -\mathcal{G} \\ \mathcal{G} & 0 \end{array}\right) $ with $\mathcal{G}=4\pi$, and the dissipative force tensor $\hat{\mathcal{D}}=\left(\begin{array}{cc}
\mathcal{D} & 0 \\ 0 & \mathcal{D} \end{array}\right)$, where 
\begin{eqnarray}
\mathcal{D}=\pi\int_{0}^{\infty}[r(\Theta')^{2}+\frac{1}{r}\sin^{2}\Theta]dr.\label{D}
\end{eqnarray}
The force $\mathbf{F}$ due to the inhomogeneous DMI strength $D$ is expressed as $\mathbf{F}=-\frac{2\gamma_{0}}{M_{s}}g_{f}{\mathcal{F}}\nabla D$. Here, we have defined the integral (Appendix C)
\begin{eqnarray}
\mathcal{F}=-\pi\int_{0}^{\infty}[r\Theta'+\sin\Theta\cos\Theta]dr,\label{ForceD}
\end{eqnarray}
and the coefficient $g_{f}=\cos\gamma$ for N\'{e}el-type skyrmion and $g_{f}=\sin\gamma$ for Bloch-type skyrmion.

When $\mathbf{m}_{p}=(\sin\theta\cos\phi,\sin\theta\sin\phi,\cos\theta)$, the force acting on the N\'{e}el-type skyrmion by the spin transfer torque is $\mathbf{F}^{s}=\cos\gamma(\sin\theta\sin\phi,-\sin\theta\cos\phi)\beta\mathcal{F}$, while the force on the Bloch-type skyrmion will be $\mathbf{F}^{s}=-\sin\gamma(\sin\theta\cos\phi,\sin\theta\sin\phi)\beta\mathcal{F}$.

The parameters in the Thiele equation (\ref{Thiele}) can be further calculated if the radial profile of the skyrmion $\Theta(r)$ is known, which then can predict the velocity of the skyrmion based on Eq.~(\ref{vx}) and (\ref{vy}). Here, we take the ansatz $\Theta(r)=2\arctan[\frac{\sinh(R/w)}{\sinh(r/w)}]$ proposed recently,\cite{CP2018} with the skyrmion radius $R=\pi D\sqrt{\frac{A}{16AK_{u}^{2}-\pi^{2}D^{2}K_{u}}}$ and the skyrmion width $w=\frac{\pi D}{4K_{u}}$.\cite{CP2018} In Fig.~\ref{Fig2}, $R$, $w$, $\mathcal{D}$, $\mathcal{F}$ are shown as the functions of DMI strength $D$ by fixing the material parameters $A=15$~pJ/m and $K_{u}=0.8$~MJ/m$^{3}$. All of them will monotonically increase when $D$ varies from $3$~mJ/m$^{2}$ to $4$~mJ/m$^{2}$. 

\section{Micromagnetic Simulations}
In this section, we study the current-driven dynamics of the magnetic skyrmion in the film by simulating the LLGS equation (\ref{LLGS}), and the simulated results are compared with the theoretical predictions from Thiele equation. The simulation parameters are set as $L=500$~nm, $W=100$~nm, $d=0.4$~nm, $M_{s}=5.8\times10^{5}$~A/m, $A=15$~pJ/m, $K_{u}=0.8$~MJ/m$^{3},\alpha=0.05$. The current amplitude is taken as $I=3\times10^{11}$~A/m, with the degree of spin polarization $p=0.4$. We restrict the simulations here to the case of N\'{e}el-type skyrmion, and set the direction of polarization $\mathbf{m}_{p}=(-1,0,0)$. The film is discretized into grids with size $1.0\times 1.0 \times 0.4$~nm$^{3}$, and the time step during the simulations is taken as $1$~fs for convergence. Three cases will be considered for the spatial modulation of DMI strength $D$ below.

\begin{figure*}[!ht]
  \centering
  \includegraphics[width=\linewidth]{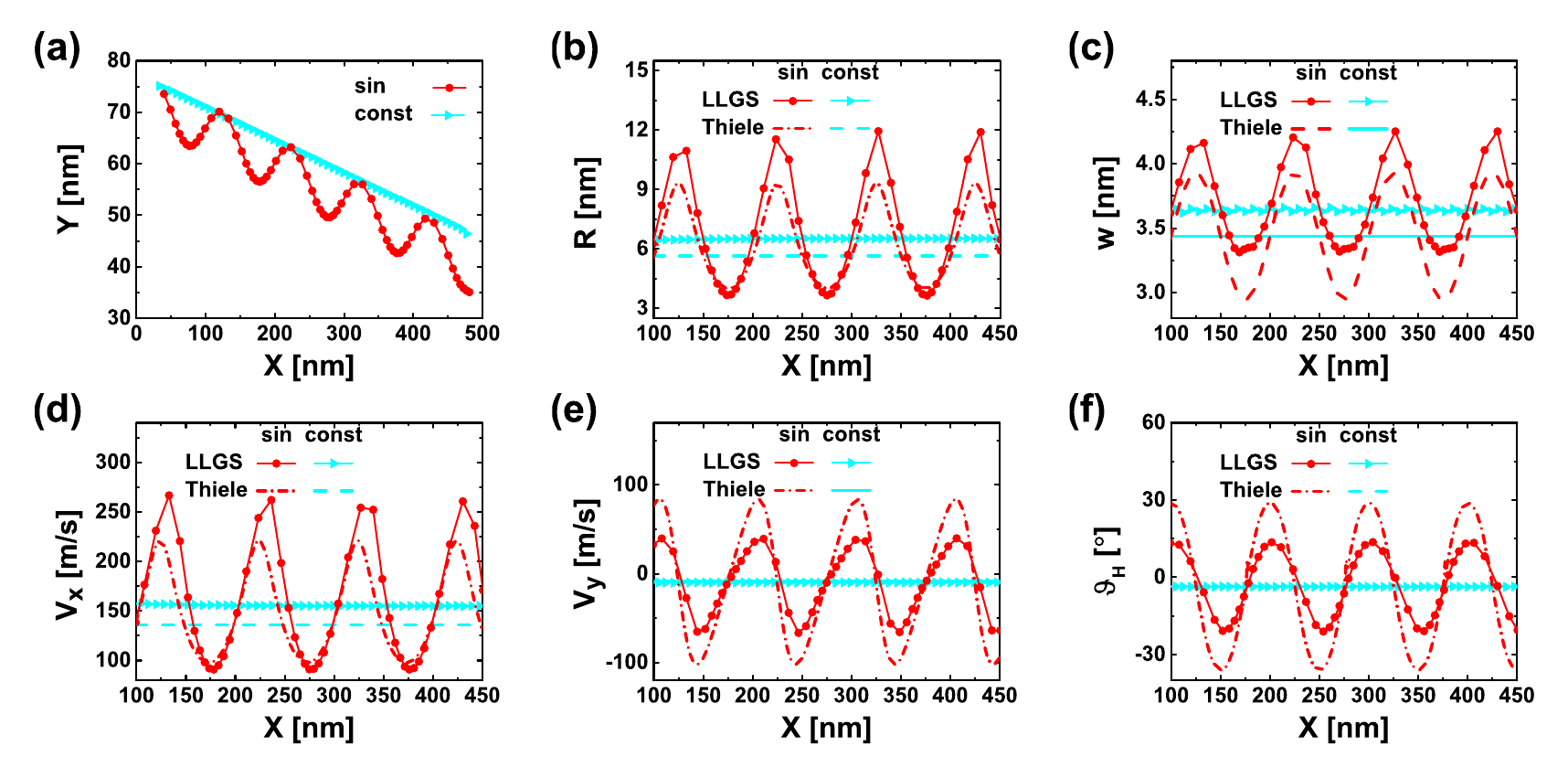}
  \caption{(Color online)(a) The trajectory of the skyrmion in the $x$-$y$ plane of the film; (b) the location-dependent skyrmion radius $R$; (c) the location-dependent skyrmion width $w$; (d)(e) the location-dependent velocity components $v_{x}$ and $v_{y}$ of the skyrmion; (f) the location-dependent Hall angle $\vartheta_{H}$ of the skyrmion. LLGS: micromagnetic simulation based on LLGS equation (\ref{LLGS}); Thiele : theoretical results based on Thiele equation. sin : $D$ is sinusoidally modulated as $3.5+0.5\sin(2\pi x/100)$~mJ/m$^{2}$ for $x\in[0,500]$~nm; const : $D$ takes constant value $3.5$~mJ/m$^{2}$ for comparison.}
  \label{Fig4}
\end{figure*}

\subsection{Linear Modulation}

We first consider that the DMI strength is linearly modulated along the $x$-direction. In Fig.~\ref{Fig3}(a), the trajectories of the skyrmion are presented when $D$ increases from 3 to 4~mJ/m$^{2}$ and decreases from 4 to 3~mJ/m$^{2}$, respectively. For comparison, the trajectory for homogeneous $D=3.5$~mJ/m$^{2}$ is also shown. The results reveal that the Hall angle $\vartheta_{H}$ can be controlled by introducing the gradient of $D$. Besides, the size of the skyrmion will also vary when $D$ is changed during the motion. In Fig.~\ref{Fig2}(b) and (c), the radius $R$ and width $w$ of the skyrmion are obtained by fitting the simulation results to the ansatz $\Theta(r)=2\arctan[\frac{\sinh(R/w)}{\sinh(r/w)}]$. We see that the fitted $R$ is generally larger than the theoretical value, which can be about $2$~nm for large $D$. On the other hand, the fitted $w$ is close to the theoretical values with a small difference about $0.3$~nm. Therefore, we can take the proposed model in Ref.~\onlinecite{CP2018} to approximately estimate the profile of the skyrmion. 

The velocity of the skyrmion is further determined from the trajectory of the skyrmion, as shown in Fig.~\ref{Fig3}(d) and (e). As $D$ is linearly increased from $3$~mJ/m$^{2}$ to $4$~mJ/m$^{2}$, both components of the velocity $v_{x}$ and $v_{y}$ will become larger. Based on the results in Section II, the force due to the linearly modulated $D$ is along the $x$-direction and is given as $\mathbf{F}_{x}=-\frac{2\gamma_{0}}{M_{s}}\mathcal{F}\nabla D$, while the force due to the spin transfer torque is along the $y$-direction with the value $\mathbf{F}_{y}^{s}=\beta\mathcal{F}$. The Thiele equation will then give the velocity
\begin{eqnarray}
v_{x}=\frac{\mathcal{G}\beta+\frac{2\gamma_{0}}{M_{s}}\alpha\nabla D\mathcal{D}}{\mathcal{G}^{2}+\alpha^{2}\mathcal{D}^{2}}\mathcal{F},\quad
v_{y}=\frac{\frac{2\gamma_{0}}{M_{s}}\mathcal{G}\nabla D-\alpha\beta\mathcal{D}}{\mathcal{G}^{2}+\alpha^{2}\mathcal{D}^{2}}\mathcal{F}.\nonumber
\end{eqnarray}
Fig.~\ref{Fig2} has shown that $\mathcal{F}$ will quickly increase from $40$ to $90$~nm during the motion, thus both of $v_{x}$ and $v_{y}$ will increase along with the DMI strength $D$. Meanwhile, $\mathcal{D}$ will also slowly increase from $15$ to $18$ (see Fig.~\ref{Fig2}(c)), which thus explains the slight increase of the magnitude of Hall angle $\vartheta_{H}=\arctan(\frac{\frac{2\gamma_{0}}{M_{s}}\mathcal{G}\nabla D-\alpha\beta\mathcal{D}}{\mathcal{G}\beta+\frac{2\gamma_{0}}{M_{s}}\alpha\nabla D\mathcal{D}})$. Similarly, when $D$ decreases from $4$~mJ/m$^{2}$ to $3$~mJ/m$^{2}$, the velocity $v_{x}$ and $v_{y}$ will both decrease, while the magnitude of the Hall angle $\vartheta_{H}$ will continuously decrease. Besides, the magnitude of Hall angle will be small for positive $\nabla D$ and be large for negative $\nabla D$ at fixed $D$, as shown in Fig.~\ref{Fig3}(a) and (f). Fig.~\ref{Fig3}(d)(e)(f) suggest that the Thiele equation can correctly capture the trend of the velocity and Hall angle with linearly modulated $D$, although some quantitative differences exist. 

\begin{figure*}[!ht]
  \centering
  \includegraphics[width=\linewidth]{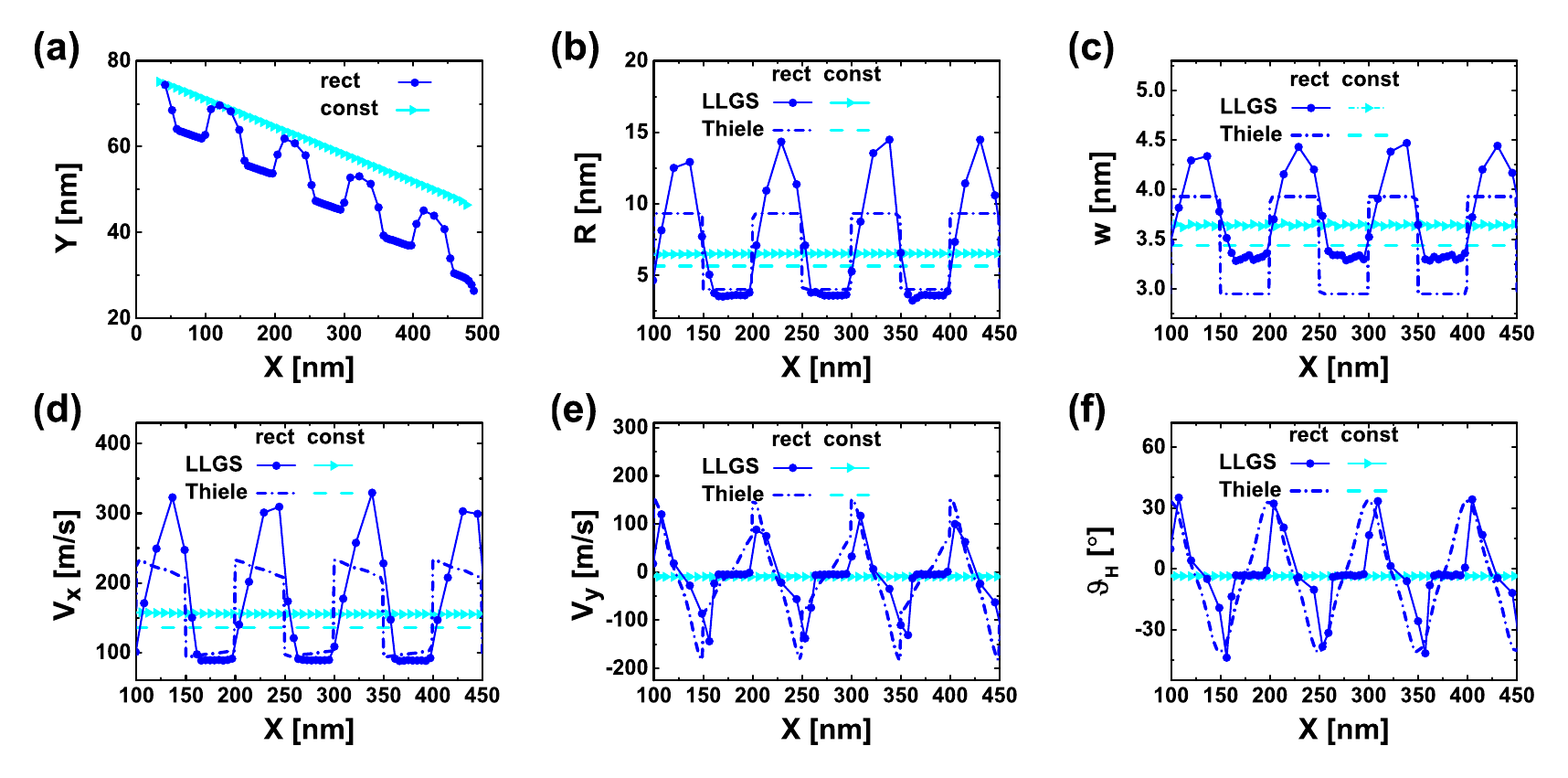}
  \caption{(Color online)(a) The trajectory of the skyrmion in the $x$-$y$ plane of the film; (b) the location-dependent skyrmion radius $R$; (c) the location-dependent skyrmion width $w$; (d)(e) the location-dependent velocity components $v_{x}$ and $v_{y}$ of the skyrmion; (f) the location-dependent Hall angle $\vartheta_{H}$ of the skyrmion. LLGS: micromagnetic simulation based on LLGS equation (\ref{LLGS}); Thiele : theoretical results based on Thiele equation. sin : $D$ takes the values $4$~mJ/m$^{2}$ or $3$~mJ/m$^2$ alternately per 50 nm for $x\in[0,500]$~nm; const : $D$ takes constant value $3.5$~mJ/m$^{2}$ for comparison.}
  \label{Fig5}
\end{figure*}

\subsection{Sinusoidal Modulation}

We further study the current-driven dynamics of skyrmion with a sinusoidal modulation of the DMI strength, where $D=3.5+0.5\sin(2\pi x/100)$ mJ/m$^{2}$ with $x\in[0,500]$~nm. Fig.~\ref{Fig4}(a) shows that the trajectory of the skyrmion will be periodically distorted by the modulated DMI. Specially, the skyrmion will be pushed away but can always move back to the trajectory if $D$ is set as $3.5$~mJ/m$^{2}$. The behavior can be understood from the periodic modulation of the Hall angle, the magnitude of which will become large for negative $\nabla D$ and vice versa. The turning points $x_{0}$ of the trajectory will then be determined by the condition $\nabla D=0$, which gives $x_{0}=25,75,\cdots,475$~nm. The skyrmion size will also be periodically modulated as expected. In Fig.~\ref{Fig4}(b) and (c), the radius $R$ and width $w$ of the skyrmion obtained by fitting the simulation results are compared to the theoretical values. The size of the skyrmion will expand and shrink periodically together with the DMI strength. In spite of quantitative differences, the theoretical calculations do can predict the variation of the skyrmion size satisfactorily. 

The velocity $\mathbf{v}$ and Hall angle $\vartheta_{H}$ have also been obtained from the micromagnetic simulations and calculated from the Thiele equation separately. As shown in Fig.~\ref{Fig4}(d)(e)(f), $\mathbf{v}$ and $\vartheta_{H}$ will oscillate around the reference values obtained for fixed $D=3.5$~mJ/m$^{2}$. Besides, $v_{x}$ will achieve it maximal value when $D$ is largest and vice versa, as determined by the factor $\mathcal{F}$ in Thiele equation. However, $v_{y}$ in this case can be either positive or negative, since $\nabla D$ here is large enough to be dominant. In the same reason, $\vartheta_{H}$ can also be positive or negative, which results in the ``push-back" trajectory given in Fig.~\ref{Fig4}(a). Moreover, the Hall angle with maximally negative value will be achieved when $\nabla D$ is negatively largest at $x=50,150,250,350,450$~nm, while the Hall angle with maximally positive value will be achieved at $x=100,200,300,400$~nm.

\subsection{Periodic Rectangular Modulation}
  
Finally, we consider the effect of the periodic rectangular modulation of DMI on the dynamics of skyrmion. In Fig.~\ref{Fig5}(a), the trajectory of skyrmion has similar ``push-back" feature, where $D$ takes the values $4$~mJ/m$^{2}$ or $3$~mJ/m$^{2}$ alternately for each $50$~nm along $x$-direction. Unlike the sinusoidal modulation case, the skyrmion size will be suddenly changed because of the abrupt increase or decrease of DMI; otherwise, the radius $R$ and width $w$ will be nearly the same in the regions with constant $D$ (see Fig.~\ref{Fig5}(b)(c)). Once again, the agreement between the theoretical and simulated results will be better when $D$ is small. 

We have assumed that the modulation of DMI should be slow in the scale of the skyrmion size in order to get the force $\mathbf{F}$ in Section II. This assumption will break down for the case here, since $D$ is a step function of the location and $\nabla D$ will give the $\delta$-pulsive force. Considering the finite size of the skyrmion, we exploit the Gaussian distribution $\frac{1}{\sqrt{\pi}\sigma}e^{-\frac{x^{2}}{\sigma^{2}}}$ to replace the $\delta(x)$ function, where $\sigma$ defines the broadening degree of the force $\mathbf{F}$. As shown in Fig.~\ref{Fig5}(d)(e)(f), the velocity $\mathbf{v}$ and Hall angle $\vartheta_{H}$ of the skyrmion are calculated from the Thiele equation by setting $\sigma=15$~nm, which can reasonably explain the simulation results. Not surprisingly, $\mathbf{v}$ and $\vartheta_{H}$ will repeatedly modulated, and the direction of $v_{y}$ will be changed to give the ``push-back" motion in Fig.~\ref{Fig5}(a). 

\section{Conclusion}
In conclusion, we have investigated the current-driven dynamics of magnetic skyrmion in chiral ferromagnetic film with spatially modulated DMI based on the Thiele equation and the micromagnetic simulations. We show that the trajectory and size of skyrmion can be manipulated by tuning the Dzyaloshinskii-Moriya interaction. Specially, the motion of the skyrmion can be sped up or slowed down by the gradient of DMI, and periodic DMI will cause the ``push-back" trajectory of the skyrmion. These phenomena could be realized by the material engineering technologies nowadays, \emph{e.g.} applying a spatially modulated gate voltage. The results here will not only advance our understanding of the rich dynamics of magnetic skyrmion further, but also can be useful to develop novel skyrmion-based spintronics devices. 

\begin{acknowledgments}
This work has been supported by NSFC Project No. 61674083 and No. 11604162 and by the Fundamental Research Funds for the Central Universities, Nankai University (Grant No. 7540).
\end{acknowledgments}

\appendix

\section{Derivation of Thiele Equation}
Here, we give the derivation of the Thiele equation from the LLGS equation (\ref{LLGS}). Taking the cross product of $\mathbf{m}$ and Eq.~(\ref{LLGS}), we will get
\begin{eqnarray}
\mathbf{m}\times\frac{d\mathbf{m}}{dt} & = & -\gamma_{0}\mathbf{m}\times(\mathbf{m}\times\mathbf{H}_{eff}) + \alpha\mathbf{m}\times(\mathbf{m}\times\frac{d\mathbf{m}}{dt})\nonumber\\
&&-\beta\mathbf{m}\times(\mathbf{m}\times(\mathbf{m}\times\mathbf{m}_{p})),\label{LLG1}
\end{eqnarray}
which can further reduce to 
\begin{eqnarray}
\mathbf{m}\times\frac{d\mathbf{m}}{dt} & = & -\gamma_{0}(\mathbf{m}\cdot\mathbf{H}_{eff})\mathbf{m}+\gamma_{0}\mathbf{H}_{eff} - \alpha\frac{d\mathbf{m}}{dt}\nonumber\\
&&+\beta\mathbf{m}\times\mathbf{m}_{p}.\label{LLG2}
\end{eqnarray}
Here, we have applied the rule of vector triple product $\mathbf{a}\times(\mathbf{b}\times\mathbf{c})=\mathbf{b}(\mathbf{a}\cdot\mathbf{c})-\mathbf{c}(\mathbf{a}\cdot\mathbf{b})$ and the relations $\mathbf{m}\cdot\mathbf{m}=1,\mathbf{m}\cdot\frac{d\mathbf{m}}{dt}=0$.
Assuming that the magnetization distribution of the skyrmion is not deformed during its motion, which is given as $\mathbf{m}(\bm{\xi})$ with the local coordinates $\bm{\xi}$ relative to its center $\mathbf{R}(t)$, then the magnetization distribution of the whole film in the lab coordinates will be $\mathbf{m}(\mathbf{r}-\mathbf{R}(t))=\mathbf{m}(\bm{\xi})$, and we will have $\frac{d\mathbf{m}}{dt}=-\bm{v}\cdot\nabla_{\bm{\xi}}\mathbf{m}$ with the skyrmion velocity $\bm{v}=\frac{d\mathbf{R}}{dt}$. Substituting $\frac{d\mathbf{m}}{dt}$ into Eq.~(\ref{LLG2}) and taking the dot product with $\nabla_{\bm{\xi}'}\mathbf{m}$, we further get
\begin{eqnarray}
&&-\nabla_{\bm{\xi}'}\mathbf{m}\cdot(\mathbf{m}\times\bm{v}\cdot\nabla_{\bm{\xi}}\mathbf{m}) = \gamma_{0}\nabla_{\bm{\xi}'}\mathbf{m}\cdot\mathbf{H}_{eff}\nonumber\\&& + \alpha(\bm{v}\cdot\nabla_{\bm{\xi}}\mathbf{m})\cdot\nabla_{\bm{\xi}'}\mathbf{m} +\beta\nabla_{\bm{\xi}'}\mathbf{m}\cdot(\mathbf{m}\times\mathbf{m}_{p}).\label{LLG3}
\end{eqnarray} 
After integrating over the whole film, Eq.~(\ref{LLG3}) will reduce to the Thiele equation as
\begin{eqnarray}
\mathcal{Q}\bm{v}-\alpha\mathcal{D}\bm{v}-\mathbf{F}-\mathbf{F}^{s}=0.\label{ThieleApp}
\end{eqnarray}
Here, $\mathcal{Q}$ is the gyromagnetic coupling tensor and $\mathcal{D}$ is the dissipative force tensor; $\mathbf{F}$ and $\mathbf{F}^{s}$ are the forces on the skyrmion due to the inhomogeneous energy density $\mathcal{E}[\mathbf{m}]$ and spin transfer torque, respectively. They are defined as the functions of the magnetization configuration $\mathbf{m}(\mathbf{r})$ as
\begin{eqnarray}
\mathcal{Q}&=&\int d\mathbf{r} (\nabla_{\bm{\xi}'}\mathbf{m}\times\nabla_{\bm{\xi}}\mathbf{m})\cdot\mathbf{m},\label{Qexp}\nonumber\\
\mathcal{D}&=&\int d\mathbf{r} \nabla_{\bm{\xi}'}\mathbf{m}\cdot\nabla_{\bm{\xi}}\mathbf{m},\label{Dexp}\nonumber\\
\mathbf{F}&=&-\frac{\gamma_{0}}{M_{s}}\int d\mathbf{r}\nabla_{\bm{\xi}'}\mathcal{E}[\mathbf{m}],\label{F}\nonumber\\
\mathbf{F}^{s}&=&\beta\int d\mathbf{r}(\nabla_{\bm{\xi}'}\mathbf{m}\times\mathbf{m})\cdot\mathbf{m}_{p}.\nonumber
\end{eqnarray}

\section{Magnetic Skyrmion Configuration}
In the polar coordinates, the magnetization configuration the skyrmion is expressed by $ \mathbf{m}(\mathbf{r})=(\sin\Theta(r)\cos\Phi(\varphi),\sin\Theta(r)\sin\Phi(\varphi),\cos\Theta(r))$. The components of $\nabla_{\bm{\xi}}\mathbf{m}$ are calculated as
\begin{eqnarray}
\frac{\partial{m}_{x}}{\partial x}&=&\Theta'\cos\Theta\cos\Phi\cos\varphi+\frac{1}{r}\sin\Theta\sin\Phi\sin\varphi,\nonumber\\
\frac{\partial{m}_{x}}{\partial y}&=&\Theta'\cos\Theta\cos\Phi\sin\varphi-\frac{1}{r}\sin\Theta\sin\Phi\cos\varphi,\nonumber\\
\frac{\partial{m}_{y}}{\partial x}&=&\Theta'\cos\Theta\sin\Phi\cos\varphi-\frac{1}{r}\sin\Theta\cos\Phi\sin\varphi,\nonumber\\
\frac{\partial{m}_{y}}{\partial y}&=&\Theta'\cos\Theta\sin\Phi\sin\varphi+\frac{1}{r}\sin\Theta\cos\Phi\cos\varphi,\nonumber\\
\frac{\partial{m}_{z}}{\partial x}&=&-\Theta'\sin\Theta\cos\varphi,\nonumber\\
\frac{\partial{m}_{z}}{\partial y}&=&-\Theta'\sin\Theta\sin\varphi.\nonumber
\end{eqnarray}
Here, we have utilized the relations 
\begin{eqnarray}
\frac{\partial r}{\partial x}=\cos\varphi,\frac{\partial r}{\partial y}=\sin\varphi,\frac{\partial \varphi}{\partial x}=-\frac{1}{r}\sin\varphi,\frac{\partial \varphi}{\partial y}=\frac{1}{r}\cos\varphi.\nonumber
\end{eqnarray}
Then the energy density $\mathcal{E}[\mathbf{m}]$ with the interfacial DMI reads
\begin{eqnarray}
\mathcal{E}[\mathbf{m}]&=&A[(\Theta')^{2}+\frac{1}{r^{2}}\sin^{2}\Theta]-K_{u}\cos^{2}\Theta\nonumber\\
&&+D(\Theta'+\frac{1}{r}\sin\Theta\cos\Theta)\cos(\Phi-\varphi).\label{ENeel}
\end{eqnarray}
The stable magnetization configuration is determined by minimizing the total energy $E[(31 Dec 2018)\Theta,\Phi]=\int d\mathbf{r}\mathcal{E}[\mathbf{m}]$ with the variational principle $\delta E=0$, which results in the following equations
\begin{eqnarray}
&&D(\Theta'+\frac{1}{r}\sin\Theta\cos\Theta)\sin(\Phi-\varphi)=0,\nonumber\\
&&\Theta''+\frac{1}{r}\Theta'-(\frac{1}{2r^{2}}+\frac{K_{u}}{2A})\sin 2\Theta+\frac{D}{Ar}\sin^{2}\Theta\cos(\Phi-\varphi)=0.\nonumber
\end{eqnarray}
Therefore, the helicity $\gamma=\Phi-\varphi$ should be $0$ or $\pi$, namely, N\'{e}el-type skyrmion. Besides, considering that $\Theta|_{r=0}=\pi$ and $\Theta|_{r\rightarrow\infty}=0$, the energy due to DMI will be minimized if $\gamma=0$ for $D>0$ or $\gamma=\pi$ for $D<0$.

Similarly, the energy density $\mathcal{E}[\mathbf{m}]$ with bulk DMI is 
\begin{eqnarray}
\mathcal{E}[\mathbf{m}]&=&A[(\Theta')^{2}+\frac{1}{r^{2}}\sin^{2}\Theta]-K_{u}\cos^{2}\Theta\nonumber\\
&&+D(\Theta'+\frac{1}{r}\sin\Theta\cos\Theta)\sin(\Phi-\varphi).\label{EBloch}
\end{eqnarray}
The variational equation $\delta\int d\mathbf{r}\mathcal{E}[\mathbf{m}]=0$ then results in the following equations
\begin{eqnarray}
&&D(\Theta'+\frac{1}{r}\sin\Theta\cos\Theta)\cos(\Phi-\varphi)=0,\nonumber\\
&&\Theta''+\frac{1}{r}\Theta'-(\frac{1}{2r^{2}}+\frac{K_{u}}{2A})\sin 2\Theta+\frac{D}{Ar}\sin^{2}\Theta\sin(\Phi-\varphi)=0.\nonumber
\end{eqnarray}
In this case, the skyrmion will be Bloch-type with the helicity $\gamma=\pi/2$ for $D>0$ or $\gamma=-\pi/2$ for $D<0$.

\section{Expressions of $\mathcal{Q},\mathcal{D},\mathbf{F},\mathbf{F}^{s}$}
We now determine the expressions of $\mathcal{Q},\mathcal{D},\mathbf{F},\mathbf{F}^{s}$ for a magnetic skyrmion in terms of the polar coordinates. First,
It is easy to see that the diagonal matrix elements $\mathcal{Q}_{xx}=\mathcal{Q}_{yy}=0$. While for its non-diagonal matrix element, straightforward calculations give $\frac{\partial\mathbf{m}}{\partial x}\times\frac{\partial\mathbf{m}}{\partial y}=\frac{1}{r}\Theta'\sin\Theta\mathbf{m}$,
therefore we can get
\begin{eqnarray}
\mathcal{Q}_{xy}=-\mathcal{Q}_{yx}=\int_{0}^{\infty}dr\int_{0}^{2\pi}d\varphi\Theta'\sin\Theta=-4\pi,\label{Qxy}
\end{eqnarray}
which is the well-known skyrmion number with a $-4\pi$ multiplier.\cite{NN2013-3}

In order to calculate $\mathcal{D}$, we first have
\begin{eqnarray}
\frac{\partial\mathbf{m}}{\partial x}\cdot\frac{\partial\mathbf{m}}{\partial x}=(\Theta')^{2}\cos^{2}\varphi+\frac{1}{r^{2}}\sin^{2}\Theta\sin^{2}\varphi,\nonumber\\
\frac{\partial\mathbf{m}}{\partial y}\cdot\frac{\partial\mathbf{m}}{\partial y}=(\Theta')^{2}\sin^{2}\varphi+\frac{1}{r^{2}}\sin^{2}\Theta\cos^{2}\varphi,\nonumber\\
\frac{\partial\mathbf{m}}{\partial x}\cdot\frac{\partial\mathbf{m}}{\partial y}=((\Theta')^{2}-\frac{1}{r^{2}}\sin^{2}\Theta)\sin\varphi\cos\varphi.\nonumber
\end{eqnarray}
Therefore, the non-diagonal matrix elements $\mathcal{D}_{xy}=\mathcal{D}_{yx}=0$ due to the integral over $\varphi$ from $0$ to $2\pi$. The diagonal matrix elements of $\mathcal{D}$ will be 
\begin{eqnarray}
\mathcal{D}_{xx}=\mathcal{D}_{yy}=\pi\int_{0}^{\infty}[r(\Theta')^{2}+\frac{1}{r}\sin^{2}\Theta]dr.\label{Dxx}
\end{eqnarray}

The force $\mathbf{F}$ can be calculated from the energy density $\mathcal{E}[\mathbf{m}]$ given in Appendix B. Since we have assumed that the skyrmion is not deformed, the only non-zero contribution to the force $\mathbf{F}$ will come from the inhomogeneous $D$, which is a function of the skyrmion center $\mathbf{R}$. Since $\nabla_{\bm{\xi}}=-\nabla_{\mathbf{R}}$, we will have
\begin{eqnarray}
\mathbf{F}=g_{f}\frac{2\pi\gamma_{0}}{M_{s}}\nabla D\int_{0}^{\infty}[r\Theta'+\sin\Theta\cos\Theta]dr.\nonumber
\end{eqnarray}
Here, the factor $g_{f}=\cos\gamma$ for N\'{e}el-type skyrmion and $g_{f}=\sin\gamma$ for Bloch-type skyrmion.

Finally, we calculate the force $\mathbf{F}^{s}$ due to the spin transfer torque. We first have
\begin{eqnarray}
\partial_{x}\mathbf{m}\times\mathbf{m}=\left(\begin{array}{c}
\Theta'\sin\Phi\cos\varphi-\frac{1}{r}\sin\Theta\cos\Theta\cos\Phi\sin\varphi\\
-\Theta'\cos\Phi\cos\varphi-\frac{1}{r}\sin\Theta\cos\Theta\sin\Phi\sin\varphi\\
\frac{1}{r}\sin^{2}\Theta\sin\varphi
\end{array}\right)\nonumber
\end{eqnarray}
\begin{eqnarray}
\partial_{y}\mathbf{m}\times\mathbf{m}=\left(\begin{array}{c}
\Theta'\sin\Phi\sin\varphi+\frac{1}{r}\sin\Theta\cos\Theta\cos\Phi\cos\varphi\\
-\Theta'\cos\Phi\sin\varphi+\frac{1}{r}\sin\Theta\cos\Theta\sin\Phi\cos\varphi\\
-\frac{1}{r}\sin^{2}\Theta\cos\varphi
\end{array}\right)\nonumber
\end{eqnarray}
For N\'{e}el-type skyrmion where $\Phi=\varphi$ or $\Phi=\varphi+\pi$, only the $y$-component of $\int d\mathbf{r}(\partial_{x}\mathbf{m}\times\mathbf{m})$ and the $x$-component of $\int d\mathbf{r}(\partial_{y}\mathbf{m}\times\mathbf{m})$ are not zero after integral over $\varphi$. While for Bloch-type skyrmion where $\Phi=\varphi\pm\frac{\pi}{2}$, only the $x$-component of $\int d\mathbf{r}(\partial_{x}\mathbf{m}\times\mathbf{m})$
and the $y$-component of $\int d\mathbf{r}(\partial_{y}\mathbf{m}\times\mathbf{m})$ are not zero after the integral. Therefore, if $\mathbf{m}_{p}=(\sin\theta\cos\phi,\sin\theta\sin\phi,\cos\theta)$, the force acting on the N\'{e}el-type skyrmion will be
\begin{eqnarray}
\mathbf{F}^{s}=\cos\gamma(\sin\theta\sin\phi,-\sin\theta\cos\phi)\beta\mathcal{F},\label{FNeel}
\end{eqnarray}
while the force acting on the Bloch-type skyrmion will be
\begin{eqnarray}
\mathbf{F}^{s}=-\sin\gamma(\sin\theta\cos\phi,\sin\theta\sin\phi)\beta\mathcal{F}.\label{FBloch}
\end{eqnarray}
Here, we have introduced the notation 
\begin{eqnarray}
\mathcal{F}=-\pi\int_{0}^{\infty}[r\Theta'+\sin\Theta\cos\Theta]dr,\nonumber
\end{eqnarray}
which has already appeared in the expression for $\mathbf{F}$.


\begin{thebibliography}{99}
\bibitem{JMMM1994} A. Bogdanov and A. Hubert, J. Magn. Magn. Mater. \textbf{138}, 255 (1994). 
\bibitem{PRL2001} A.N. Bogdanov and U.K. R\"{o}$\beta$ler, Phys. Rev. Lett. \textbf{87}, 037203 (2001).
\bibitem{Nat2006} U.K. R\"{o}$\beta$ler, A.N. Bogdanov, and C. Pfleiderer, Nature, \textbf{442}, 797(2006).
\bibitem{Sci2009} S. M\"{u}hlbauer, B. Binz, F. Jonietz, C. Pfleiderer, A. Rosch, A. Neubauer, R. Georgii, and P. B\"{o}ni, Science, \textbf{323}, 915 (2009).
\bibitem{Nat2010} X.Z. Yu, Y. Onose, N. Kanazawa, J.H. Park, J.H. Han, Y. Matsui, N. Nagaosa, and Y. Tokura, Nature, \textbf{465}, 901 (2010).
\bibitem{NM2010} X.Z. Yu, N. Kanazawa, Y. Onose, K. Kimoto, W. Z. Zhang, S. Ishiwata, Y. Matsui, and Y. Tokura, Nature Mater. \textbf{10}, 106 (2010).
\bibitem{NP2011} S. Heinze, K. von Bergmann, M. Menzel, J. Brede, A. Kubetzka, R. Wiesendanger, G. Bihlmayer, and S. Bl\"{u}gel, Nat. Phys. \textbf{7} 713 (2011).
\bibitem{NP2012} T. Schulz, R. Ritz, A. Bauer, M. Halder, M. Wagner, C. Franz, C. Pfleiderer, K. Everschor, M. Garst, and A. Rosch, Nat. Phys. \textbf{8}, 301 (2012).   
\bibitem{PRL2009} A. Neubauer, C. Pfleiderer, B. Binz, A. Rosch, R. Ritz, P.G. Niklowitz, and P. Boni, Phys. Rev. Lett. \textbf{102}, 186602 (2009).
\bibitem{PRL2011-1} N. Kanazawa, Y. Onose, T. Arima, D. Okuyama, K. Ohoyama, S. Wakimoto, K. Kakurai, S. Ishiwata, and Y. Tokura, Phys. Rev. Lett. \textbf{106}, 156603 (2011).
\bibitem{PRL2011-2} J.D. Zang, M. Mostovoy, J.H. Han, and N. Nagaosa, Phys. Rev. Lett. \textbf{107}, 136804 (2011).
\bibitem{Sci2013} N. Romming, C. Hanneken, M. Menzel, J. E. Bickel, B. Wolter, K. von Bergmann, A. Kubetzka, and R. Wiesendanger, Science \textbf{341},636 (2013).
\bibitem{NN2013-1} J. Sampaio, V. Cros, S. Rohart, A. Thiaville, and A. Fert, Nat. Nanotechnol. \textbf{8}, 839 (2013).
\bibitem{NN2013-2} J. Iwasaki, M. Mochizuki, and N. Nagaosa, Nat. Nanotechnol. \textbf{8}, 742 (2013).
\bibitem{PRL2013} Y.F. Li, N. Kanazawa, X.Z. Yu, A. Tsukazaki, M. Kawasaki, M. Ichikawa, X.F. Jin, F. Kagawa, and Y. Tokura, Phys. Rev. Lett. \textbf{110}, 117202 (2013).
\bibitem{NC2014} Y. Zhou and M. Ezawa, Nat. Commun. \textbf{5}, 4652 (2014).
\bibitem{Sci2015} W. Jiang, P. Upadhyaya, W. Zhang, G. Yu, M. B. Jungfleisch, F. Y. Fradin, J. E. Pearson, Y. Tserkovnyak,
K. L. Wang, O. Heinonen, S. G. E. te Velthuis, and A. Hoffmann, Science \textbf{349}, 283 (2015).
\bibitem{NP2017-1} W.J. Jiang, X.C. Zhang, G.Q. Yu, W. Zhang, X. Wang, M.B. Jungfleisch, J.E. Pearson, X.M. Cheng, O.Heinonen, K.L. Wang, Y. Zhou, A. Hoffmann, and S.G.E. te Velthuis, Nat. Phys. \textbf{13}, 162 (2017).
\bibitem{NP2017-2} K. Litzius, I. Lemesh, B. Kruger, P. Bassirian, L. Caretta, K. Richter, F. Buttner, K. Sato, O.A. Tretiakov, J. Forster, R.M. Reeve, M. Weigand, L. Bykova, H. Stoll, G. Schutz, G.S.D. Beach, and M. Klaui, Nat. Phys. \textbf{13}, 170 (2017).
\bibitem{NN2017} P.-J. Hsu, A. Kubetzka, A. Finco, N. Romming, K. von Bergmann, and R. Wiesendanger, Nat. Nanotechnol. \textbf{12}, 123 (2017).
\bibitem{NN2013-3} N. Nagaosa and Y. Tokura, Nat. Nanotechnol. \textbf{8}, 899 (2013).
\bibitem{NRM1} R. Wiesendanger, Nat. Rev. Mater. \textbf{1}, 16044 (2016).
\bibitem{NRM2} A. Fert, N. Reyren, and V. Cros, Nat. Rev. Mater. \textbf{2}, 17031 (2017).
\bibitem{Proc} W. Kang, Y. Huang, X.C. Zhang, Y. Zhou, and W. Zhao, Proc. IEEE \textbf{104}, 2040 (2016).
\bibitem{NatEle2018} S. Woo, K. M. Song, X. Zhang, M. Ezawa, Y. Zhou, X. Liu, M. Weigand, S. Finizio, J. Raabe, M.-C. Park, K.-Y. Lee, J. W. Choi, B.-C. Min, H. C. Koo, and J. Chang, Nat. Electron. \textbf{1}, 288 (2018).
\bibitem{Dzya} I. Dzyaloshinskii, J. Phys. Chem. Solids \textbf{4}, 241 (1958).
\bibitem{Moriya} T. Moriya, Phys. Rev. \textbf{120}, 91 (1960).
\bibitem{PRB2015} S.-A. Siegfried, E. V. Altynbaev, N. M. Chubova, V. Dyadkin, D. Chernyshov, E. V. Moskvin, D. Menzel, A. Heinemann, A. Schreyer, and S. V. Grigoriev, Phys. Rev. B \textbf{91}, 184406 (2015).
\bibitem{SR2015} T. Koretsune, N. Nagaosa, and R. Arita, Sci. Rep. \textbf{5}, 13302 (2015).
\bibitem{PRB2016} X. Ma, G. Yu, X. Li, T. Wang, D. Wu, K.S. Olsson, Z. Chu, K. An, J.Q. Xiao, K.L. Wang, and X. Li, Phys. Rev. B \textbf{94}, 180408(R) (2016).
\bibitem{SR2016} A. Belabbes, G. Bihlmayer, S. Bl\"{u}gel, and A. Manchon, Sci. Rep. \textbf{6}, 24634 (2016).
\bibitem{PRL2017-1} A.L. Balk, K-W. Kim, D.T. Pierce, M.D. Stiles, J. Unguris, S.M. Stavis, \textbf{119}, 077205 (2017).
\bibitem{PRL2017-2} G. Beutier, S. P. Collins, O. V. Dimitrova, V. E. Dmitrienko, M.I. Katsnelson, Y.O. Kvashnin, A.I. Lichtenstein, V. V. Mazurenko, A.G.A. Nisbet, E. N. Ovchinnikova, and D. Pincini, Phys. Rev. Lett. \textbf{119}, 167201 (2017).
\bibitem{NL2018} T. Srivastava, M. Schott, R. Juge, V. K\v{r}i\v{z}akov\'{a}, M. Belmeguenai, Y. Roussign\'{e}, A. Bernand-Mantel, L. Ranno, S. Pizzini, S.-M. Che r\'{i}f, A. Stashkevich, S. Auffret, O. Boulle, G. Gaudin, M. Chshiev, C. Baraduc, and H. B\'{e}a, Nano. Lett. \textbf{18}, 4871 (2018).
\bibitem{PRB2018} J. Suwardy, K. Nawaoka, J. Cho, M. Goto, Y. Suzuki, and S. Miwa, Phys. Rev. B \textbf{98}, 144432 (2018).
\bibitem{Nanoscale2018} A. Cao, X. Zhang, B. Koopmans, S. Peng, Y. Zhang, Z. Wang, S. Yan, H. Yang, and W. Zhao, Nanoscale \textbf{10}, 12062 (2018).
\bibitem{SR2018} H. Yang, O. Boulle, V. Cros, A. Fert, and M. Chshiev, Sci. Rep. \textbf{8}, 12356 (2018).
\bibitem{Book} S. Seki and M. Mochizuki, \emph{Skyrmions in Magnetic Materials} (Springer, Switzerland, 2016).
\bibitem{CP2018} X.S. Wang, H.Y. Yuan, and X.R. Wang, Commun. Phys. \textbf{1}, 31 (2018).
\end{thebibliography}
\end{document}